\RequirePackage{fix-cm}
\RequirePackage{fixltx2e}
\documentclass[12pt,twoside,english,aip, jcp,floatfix]{revtex4-1}

\usepackage[T1]{fontenc}
\usepackage[latin9]{inputenc}
\usepackage{geometry}
\usepackage{xcolor}
\usepackage{babel}
\usepackage{rotating}
\usepackage{amsmath}
\usepackage{amssymb}
\usepackage{graphicx}
\usepackage[unicode=true,pdfusetitle,
 bookmarks=false,
 breaklinks=true,pdfborder={0 0 0},pdfborderstyle={},backref=false,colorlinks=true]
 {hyperref}
\hypersetup{
 citecolor = blue, linkcolor = blue,  urlcolor  = blue}

\makeatletter

\providecommand{\tabularnewline}{\\}

\usepackage{orcidlink}

\makeatother

\begin{document}

\title{Comparison of recent estimators of uncertainty on the mean for small
measurement samples with normal and non-normal error distributions}

\author{Pascal PERNOT \orcidlink{0000-0001-8586-6222}}

\affiliation{Institut de Chimie Physique, UMR8000 CNRS,~\\
Université Paris-Saclay, 91405 Orsay, France}
\email{pascal.pernot@cnrs.fr}

\author{Jean-Paul BERTHET}

\affiliation{C2RMF, Palais du Louvre, 14 quai François Mitterand, 75001 Paris,
France}
\email{jean-paul.berthet@cnrs.fr}

\begin{abstract}
\noindent We review the alternative proposals introduced recently
in the literature to update the standard formula to estimate the uncertainty
on the mean of repeated measurements, and we compare their performances
on synthetic examples with normal and non-normal error distributions.
\end{abstract}
\maketitle

\section{Introduction}

There is a running debate on the proper way to treat Type A uncertainty
in a revised version of the GUM,\citep{GUM} aimed at reducing the
discrepancy between residual frequentist and prominent Bayesian sections
of the document and its supplements, notably Supplement 1 about the
Monte Carlo method.\citep{GUMSupp1}

Let us consider a sample of measurements $\{x_{i};\ i=1,N\}$ for
which the measurement uncertainty $\sigma$ is unknown, and let us
note $s$ the sample standard deviation ($s=\sqrt{1/(N-1)\sum_{i=1}^{N}(x_{i}-\bar{x})}$).
The current version of the GUM recommends the frequentist approach
to estimate the uncertainty on the (arithmetic) sample mean $\bar{x}$,
as $u_{fr}=s/\sqrt{N}$. The proposed Bayesian revision of the GUM
implements a model using a so-called \emph{non-informative} prior
(NIP) distribution on $\sigma$, which results in a new estimator
$u_{NIP}=\sqrt{(N-1)/(N-3)}\thinspace u_{fr}$. The multiplicative
factor derives from the standard deviation of a Student's-\emph{t}
distribution with $\nu=N-1$ degrees of freedom.\citep{Kacker2003}

This Bayesian formula has two major impacts on the measurement daily
practice: (i) it enlarges considerably the uncertainty for small samples;
and (ii) it prevents the use of samples with $N=2$ or 3. Both features
are seen as problematic by many actors in the field and several propositions
have been made in the past years to overcome them.

On one hand, several authors are fighting the Bayesian approach and
propose either to stick to the previous version or to replace it with
another, non-Bayesian, one. On this side, for instance, Huang\citep{Huang2020}
favors the use of an \emph{unbiased} estimator of standard deviation,
note $u_{fru}$ (the standard estimator $u_{fr}$ is based on an unbiased
estimator of the variance).

On the other hand, propositions have been made to amend the non-informative
Bayesian solution. Kacker introduced \emph{ad hoc} correction factors
for samples with $N=2,3$.\citep{Kacker2003} This however does not
mitigates the excessive enlargement problem, and several authors proposed
to replace the non-informative prior with informative ones. Recently,
Cox and Shirono\citep{Cox2017} provided a formula resulting from
the introduction of bounds to the non-informative Jeffrey's prior.
Then, O'Hagan and Cox\citep{Ohagan2021} introduced two new informative
priors for metrology, making use of the \emph{a priori} knowledge
of an expert on the measurement process. These are labeled as \emph{mildly}
informative (MIP) and \emph{strongly} informative (SIP) priors. Finally
(as of today), Cox and O'Hagan\citep{Cox2022} proposed a strong departure
from the standard scheme by replacing the mean value and standard
deviation by the \emph{median} and a newly defined \emph{characteristic
uncertainty}, which is the half of the half-width of a 95\% probability
interval.

It has to be noted that several of the mentioned approaches make an
implicit use of the normality hypothesis of the measurement errors.
This is the case notably of the unbiased standard deviation estimator
$u_{fru}$ and of the Bayesian estimators that are using a Gaussian
likelihood function. There is no evidence that this is an acceptable
hypothesis for scientific measurements. For instance, Bailey\citep{Bailey2017}
showed that even for high accuracy measurements, a Student's-$t$
distribution with a small number of degrees of freedom (3 or 4) would
be more appropriate.

The aim of the present study is twofold: (1) to compare the properties
of these concurrent uncertainty estimators on synthetic samples of
measurements; and (2) to assess the robustness of these estimators
to perturbations of the normal measurement errors paradigm. To the
best of our knowledge, such a direct comparison is missing from the
literature, and the impact of non-normal error distributions is generally
ignored in the existing studies.

Sect.\,\ref{sec:Simulation-setup} introduces the estimators and
error distributions used for the Monte Carlo simulations. The outcomes
of these simulations are presented and analyzed in Sect.\,\ref{sec:Results-and-Discussion}.
Our findings are summarized in Sect.\,\ref{sec:Conclusion}.

\section{The simulation setup\label{sec:Simulation-setup}}

The properties of the uncertainty estimation models are assessed by
Monte Carlo simulations, where random error samples with prescribed
distribution are drawn multiple times and used as measurement data.
Monte Carlo samples of uncertainties are generated for all estimators
and error distributions described below using the \texttt{R} language.\citep{RTeam2019}

\subsection{Uncertainty estimation models}

The following models have been extracted from the recent studies presented
in the Introduction. They are straightforwardly tagged as non-Bayesian
or Bayesian, although the choice for the \emph{characteristic} uncertainty
estimator is guided by the choice of a probability interval estimation
method.
\begin{itemize}
\item Non-Bayesian estimators
\begin{itemize}
\item $u_{fr}=s/\sqrt{N}$ where $s$ is the sample standard deviation.
This is the frequentist model used in the present version of the GUM.\citep{GUM}
$u_{fr}$ is the square root of an \emph{unbiased} \emph{variance}
estimator.
\item $u_{fru}=u_{fr}/c_{4}$ with $c_{4}=\sqrt{2/(N-1)}\ \Gamma(N/2)/\Gamma((N-1)/2)$
is an \emph{unbiased standard deviation} estimator advocated by Huang.\citep{Huang2020}
A simplified version replacing $c_{4}$ by $c_{4}^{*}=\sqrt{(N-1.5)/(N-1)}$
(to avoid the computation of the gamma function) has been provided
by Brugger.\citep{Brugger1969}
\item $u_{frun}=u_{fr}\,\sqrt{(N-1)/[N-1.5-(\kappa-3)/4]}$, where $\kappa$
is the distribution \emph{kurtosis}, is an approximate \emph{unbiased
standard deviation estimator for non-normal distributions}, adapting
Brugger's formula to account for excess kurtosis.\citep{Wikipedia1}
Note that for small values of $N$, $\kappa$ cannot be reasonably
estimated from the data, and an hypothesis on the error distribution
is necessary.
\end{itemize}
\item Bayesian estimators 
\begin{itemize}
\item $u_{[N|M|S]IP}=\sqrt{d^{*}/(d^{*}-2)\,v^{*}/N}$ regroups the \emph{non-informative}
prior (NIP)\citep{Kacker2003} and the mild and strong informative
priors (MIP and SIP) proposed by O'Hagan and Cox.\citep{Ohagan2021}
The parameters $d^{*}$ and $v^{*}$ depend on the model:
\begin{itemize}
\item NIP: $d^{*}=N-1$ and $v^{*}=s^{2}$ 
\item MIP: $d^{*}=N+2$ and $v^{*}=(3v+(N-1)*s^{2})/(N+2)$ 
\item SIP: $d^{*}=N+7$ and $v^{*}=(8v+(N-1)*s^{2})/(N+7)$ 
\end{itemize}
where $v$ is the \emph{a priori} value of the \emph{variance}. The
MIP simulates the inclusion of a pseudo-sample of four measurements
with variance $v$ to the analyzed data, while this amounts to nine
pseudo-measurements for SIP. 
\item $u_{Cox}=\phi u_{fr}$ has been proposed by Cox and Shirono.\citep{Cox2017}
$\phi$ is given by Eq. 14 in the original article and depends on
the limits $\sigma_{min}$ and $\sigma_{max}$. We consider here that
they are given by $\sigma_{min}^{2}=v/3$ and $\sigma_{max}^{2}=3v$,
where $v$ is an a priori value of the variance. This choice is expected
to nearly correspond to the SIP prior.\citep{Cox2022} 
\item $u_{char}=k_{N-1}u_{fr}/2$, where $k_{\nu}$ is the 97.5\,\% quantile
of the Student's-\emph{t} distribution with $\nu$ degrees of freedom,
is the \emph{characteristic} uncertainty proposed by Cox and O'Hagan.\citep{Cox2022}
It is possible to define characteristic uncertainties for all the
types of priors presented above.\citep{Ohagan2021} The present one
is based on the NIP version. 
\end{itemize}
\end{itemize}

\subsection{Test distributions}

A set of four \emph{zero-centered}, \emph{unit-variance} distributions
with different kurtosis values ($\kappa$) were selected for the analysis:
\begin{itemize}
\item \emph{Unifu}: uniform between $\pm\sqrt{3}$ ($\kappa=9/5$) 
\item \emph{Norm}: standard Normal ($\kappa=3$) 
\item \emph{Laplu}: standard Laplace (symmetric exponential), scaled by
$1/\sqrt{2}$ ($\kappa=6$) 
\item \emph{T3u}: Students-$t(\nu=3)$, scaled by $1/\sqrt{3}$ ($\kappa=\infty$) 
\end{itemize}
We did not consider asymmetric distributions, as the treatment of
asymmetric uncertainties is beyond the goal of this short study.

\section{Results and Discussion\label{sec:Results-and-Discussion}}

As some estimators do not handle less than four measurements, this
was used as our smallest test sample size ($N=4$). The true value
in this case is $u_{true}=1/\sqrt{N}=0.5$. A control example with
$N=40$ ($u_{true}=0.158$) was also used to assess the convergence
of the estimators with increasing $N$ values. 

Monte Carlo samples ($M=10^{4}$) of the mean of $N$ measurements
$\mu_{N}$ and its uncertainty are generated for both $N$ values,
and for all uncertainty estimators and error distributions. Summary
statistics described below are derived from these samples. Note that
the $u_{Cox}$ method occasionally returns numerical exceptions, which
are filtered out from the Monte Carlo results.

\subsection{MC samples validation}

In order to validate the chosen distributions and the sampling procedure,
we estimate the mean, standard deviation and kurtosis of the $M$
generated sample means $\mu_{N}$ for all distributions. The values
are reported in Table\,\ref{tab:Table1}. Except for the mean, the
statistical uncertainties are obtained by bootstrapping\citep{Efron1979,Efron1991}
with 5000 repeats. 
\begin{table}[t]
\noindent \begin{centering}
\begin{tabular}[t]{rlr@{\extracolsep{0pt}.}lr@{\extracolsep{0pt}.}lr@{\extracolsep{0pt}.}lr@{\extracolsep{0pt}.}lr@{\extracolsep{0pt}.}lr@{\extracolsep{0pt}.}l}
\hline 
$N$ & Statistic & \multicolumn{2}{c}{Norm } & \multicolumn{2}{c}{Unifu } & \multicolumn{2}{c}{Laplu } & \multicolumn{2}{c}{T3u} & \multicolumn{2}{c}{} & \multicolumn{2}{c}{Target}\tabularnewline
\hline 
4 & MC mean & -0&006(5)  & 0&008(5)  & 0&002(5) & 0&005(5)  & \multicolumn{2}{c}{} & 0&0\tabularnewline
 & MC s.d. & 0&499(4)  & 0&501(3)  & 0&503(5) & 0&50(1)  & \multicolumn{2}{c}{} & 0&5\tabularnewline
 & MC kurtosis & 3&10(5)  & 2&70(3)  & 4&2(2) & \multicolumn{2}{c}{29(1) } & \multicolumn{2}{c}{} & \multicolumn{2}{c}{-}\tabularnewline
\noalign{\vskip\doublerulesep}
40 & MC mean & -0&001(2)  & -0&000(2)  & 0&001(2) & -0&001(2)  & \multicolumn{2}{c}{} & 0&0\tabularnewline
 & MC s.d. & 0&158(1)  & 0&157(1)  & 0&159(1) & 0&159(2)  & \multicolumn{2}{c}{} & 0&158\tabularnewline
 & MC kurtosis & 3&03(5)  & 2&98(4)  & 3&18(6) & 5&6(8)  & \multicolumn{2}{c}{} & \multicolumn{2}{c}{-}\tabularnewline
\hline 
\end{tabular}
\par\end{centering}
\caption{\label{tab:Table1}Statistical summaries of MC sample for all distributions.}
\end{table}

The generated samples are conform with the expected properties, i.e.
a mean equal to 0 (within statistical errors) and a standard deviation
equal to $u_{true}$. For the kurtosis, one sees that the distribution
of mean values is not necessarily normal, although all the kurtosis
values are closer to 3 than those of the corresponding error distributions.
For the larger measurement sample size, $N=40$, one expects the distribution
of the means to converge to a normal distribution independently of
the error distribution (Central Limit Theorem), which is assessed
by the kurtosis values getting closer to 3. Note that the samples
generated from the Student's distribution are still far from normally
distributed. 

\subsection{Comparison of uncertainty estimators for the normal distribution
of errors}

\begin{figure}[t]
\noindent \begin{centering}
\includegraphics[height=8cm]{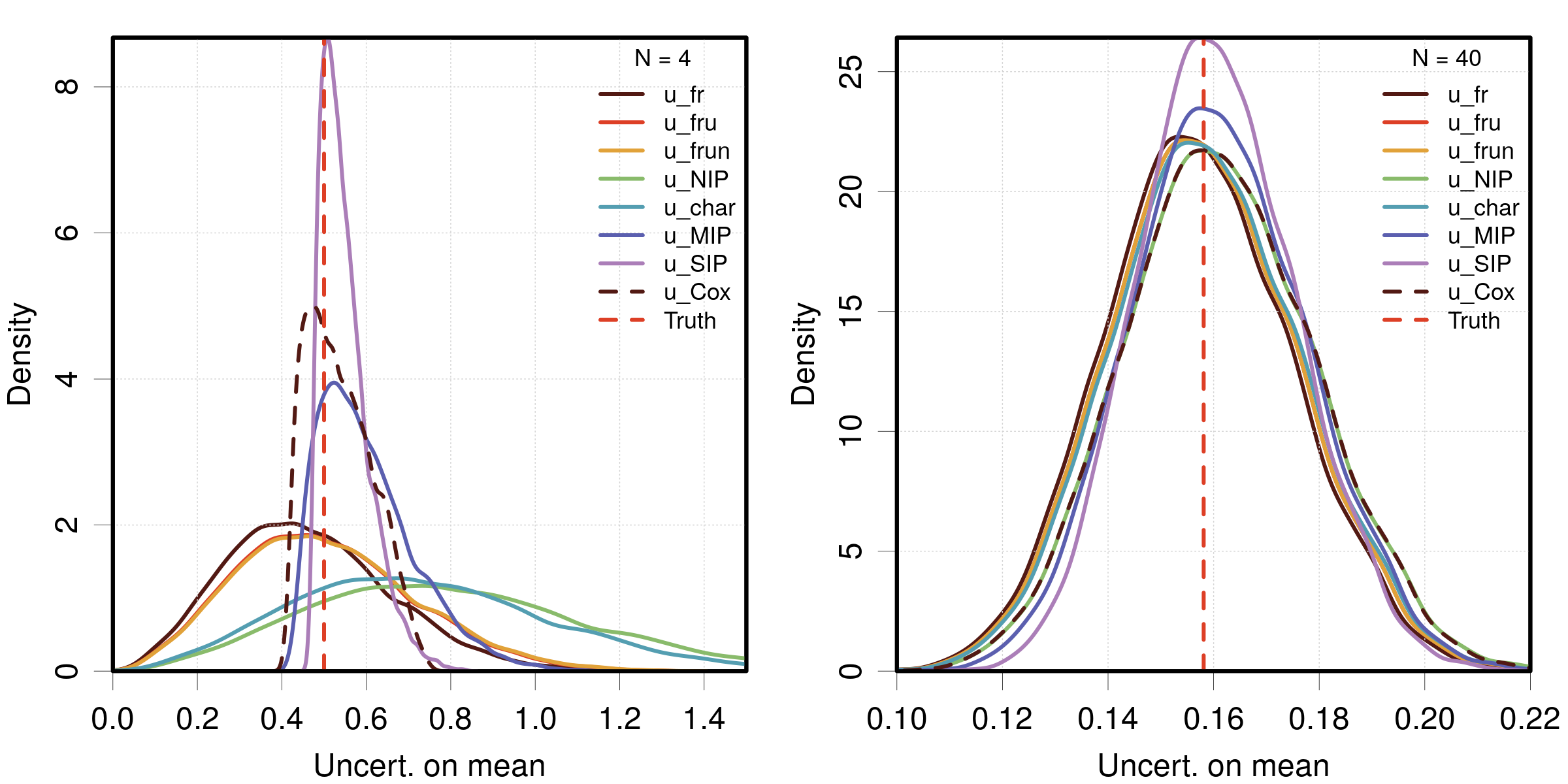}
\par\end{centering}
\caption{\label{fig:PDFs}Density plots of uncertainty Monte Carlo samples
for all uncertainty estimators and two measurement counts from a normal
distribution: (left) $N=4$; (right) $N=40$.}
\end{figure}

Let us now focus on the Normal errors model, for which most of the
uncertainty estimators are designed. Fig. \ref{fig:PDFs} reports
the probability density functions (pdf) of the Monte Carlo samples
of \emph{uncertainties} $u_{X}$ for all uncertainty estimators. 

For $N=4$, one sees clearly three groups: $u_{fr}$, $u_{fru}$ and
$u_{frun}$; $u_{NIP}$ and its derivative $u_{char}$; and the three
Bayesian models with informative priors. All distributions are strongly
skewed, which is expected at least for $u_{fr}$, as for a normal
error distribution ${\displaystyle \sqrt{N-1}s/u_{true}}$ has a chi
distribution with $N-1$ degrees of freedom. 

Another salient feature is that the distributions for the Bayesian
estimators with informative priors are much more concentrated around
the true value than all the other ones. In consequence, there is for
these estimators a much lower risk to predict a strongly under- or
over-estimated uncertainty.

For $N=40$, all density curves become closely packed, but one might
still discern two groups, namely $u_{fr}$, $u_{fru}$, $u_{NIP}$
and $u_{char}$, with a mode to the left of the true value, and $u_{MIP}$,
$u_{SIP}$ and $u_{Cox}$, with their mode closer to the true value.
All distributions are still slightly skewed. 

The skewness of the distributions at small $N$ raises the problem
of the choice of a location statistic for the uncertainty. We consider
three of them: arithmetic mean $\left\langle u_{X}\right\rangle $;
median of $u_{X}$; and root mean squared uncertainty $\sqrt{\left\langle u_{X}^{2}\right\rangle }$
(which is often used to average standard deviations). The robust Inter-Quartile
Range (IQR) is used to measure the scale of the distribution. We consider
also two probabilities to complement these statistics: 
\begin{itemize}
\item the probability to get an underestimation of uncertainty
\begin{equation}
P_{<}=P(u_{X}<u_{true})=\left\langle \boldsymbol{1}(u_{X}<u_{true})\right\rangle 
\end{equation}
where $\boldsymbol{1}(x)$ is the indicator function. The overestimation
probability is the complement to 1.\textcolor{orange}{{} }
\item the probability to lie within 20\% of the true value
\begin{equation}
P_{20}=P(u_{true}/1.2<u_{X}<1.2\times u_{true})
\end{equation}
which somewhat relaxes the difficulty of choosing a pertinent location
statistic.
\end{itemize}
Table \ref{tab:Table2-2} and Figs.\ref{fig:Statistics4}-\ref{fig:Statistics40}
report these statistics for all uncertainty estimators. 

\begin{table}[t]
\noindent \begin{centering}
\begin{turn}{90}
\begin{tabular}{llllllllllll}
\hline 
$N$ & Statistic & $u_{fr}$ & $u_{fru}$ & $u_{frun}$ & $u_{NIP}$ & $u_{char}$ & $u_{MIP}$ & $u_{SIP}$ & $u_{Cox}$ &  & Target\tabularnewline
\hline 
4 & $\left\langle u_{X}\right\rangle $ & 0.463(2)  & \textbf{0.503(2) } & 0.508(2)  & 0.803(3)  & 0.738(3)  & 0.603(1)  & 0.5504(6)  & 0.5378(8) &  & 0.5\tabularnewline
 & $\mathrm{med}(u_{X})$ & 0.446(2)  & 0.484(3)  & 0.488(3)\textbf{ } & 0.772(4)  & 0.709(4)  & 0.580(1)  & 0.5370(6)  & 0.527(1) &  & 0.5\tabularnewline
 & $\sqrt{\left\langle u^{2}(x)\right\rangle }$ & \textbf{0.502(2)}  & 0.545(2)  & 0.550(2)  & 0.870(3)  & 0.799(3)  & 0.614(1)  & 0.5535(6)  & 0.5432(8) &  & 0.5\tabularnewline
 & IQR & 0.266(3)  & 0.288(3)  & 0.291(3)  & 0.460(5)  & 0.423(5)  & 0.153(2)  & \textbf{0.0740(9) } & 0.120(1) &  & small\tabularnewline
 & $P_{<}$ & 0.601(5)  & 0.529(5)  & 0.522(5)\textbf{ } & 0.191(4)  & 0.237(4)  & 0.191(4)  & 0.191(4)  & 0.377(5) &  & 0.5\tabularnewline
 & $P_{20}$ & 0.331(5)  & 0.324(5)  & 0.324(5)  & 0.178(4)  & 0.209(4)  & 0.564(5)  & \textbf{0.824(4) } & 0.772(4) &  & high\tabularnewline
\noalign{\vskip\doublerulesep}
40 & $\left\langle u_{X}\right\rangle $ & 0.1570(2)  & \textbf{0.1580(2) } & \textbf{0.1580(2) } & 0.1612(2)  & 0.1588(2)  & 0.1610(2)  & 0.1608(1)  & 0.1611(2) &  & 0.158\tabularnewline
 & $\mathrm{med}(u_{X})$ & 0.1565(2)  & \textbf{0.1576(2) } & \textbf{0.1576(2)}  & 0.1607(2)  & \textbf{0.1583(2)}  & 0.1605(2)  & 0.1603(2)  & 0.1607(2) &  & 0.158\tabularnewline
 & $\sqrt{\left\langle u^{2}(x)\right\rangle }$ & \textbf{0.1580(2)}  & 0.1590(2)  & 0.1590(2)  & 0.1622(2)  & 0.1598(2)  & 0.1619(2)  & 0.1615(2)  & 0.1621(2) &  & 0.158\tabularnewline
 & IQR & 0.0241(3)  & 0.0243(3)  & 0.0243(3)  & 0.0248(3)  & 0.0244(3)  & 0.0229(3)  & \textbf{0.0204(3) } & 0.0248(3) &  & small\tabularnewline
 & $P_{<}$ & 0.537(5)  & \textbf{0.513(5) } & \textbf{0.513(5) } & 0.446(5)  & \textbf{0.495(5)}  & 0.446(5)  & 0.446(5)  & 0.446(5) &  & 0.5\tabularnewline
 & $P_{20}$ & 0.888(3)  & 0.890(3)  & 0.890(3)  & 0.888(3)  & 0.891(3)  & 0.916(3)  & \textbf{0.950(2) } & 0.889(3) &  & high\tabularnewline
\hline 
\end{tabular}
\end{turn}
\par\end{centering}
\caption{\label{tab:Table2-2}Statistics for Monte Carlo samples uncertainty
estimators. Except for the arithmetic mean, the estimation uncertainties
(in parenthesis notation) are obtained by bootstrapping\citep{Efron1979,Efron1991}
with 5000 repeats. For each statistic, the values compatible with
the target (last column) within $\pm2\sigma$ are noted in bold. In
absence of a target, the smallest or largest values are emphasized.}
\end{table}

Let us first consider the small sample ($N=4$). It is striking that
the $u_{fru}$ method gives the best estimate using the mean, while
the standard $u_{fr}$ gives the best estimate using the root mean
squared statistic. This is consistent with their properties, $u_{fr}$
resulting from an unbiased estimation of the variance ($u_{fr}^{2}$),
while $u_{fru}$ is an unbiased estimator of standard deviation. Globally,
the median does not provide an improvement over the other metrics.
One can also note that for the normal distribution, $u_{frun}$ is
practically indistinguishable from $u_{fru}$, a credit to Brugger's
formula.\citep{Brugger1969}

Based on the location statistics, all Bayesian estimators overestimate
the uncertainty, albeit to a lesser extent for those based on an informative
prior. Although they are considered to encode the same information
level,\citep{Cox2022} $u_{Cox}$ performs slightly better than $u_{SIP}$
for the three location estimators. 

If one considers the inter-quartile range, $u_{SIP}$ provides the
most concentrated distribution, followed by $u_{Cox}$. The largest
spread is observed for $u_{NIP}$.

Among the frequentist methods, $u_{fr}$ has the highest risk to \emph{under}-estimate
the uncertainty($P_{<}=0.62$), while $u_{fru}$ is closer to a balanced
estimation ($P_{<}=0.53$). As noted from the location statistics,
all Bayesian methods tend to \emph{over}-estimate the uncertainty,
with probabilities between $1-P_{<}=0.79$ for $u_{NIP}$ and $1-P_{<}=0.60$
for $u_{Cox}$. 

Finally, the probability to be within 20\% of the true value, $P_{20}$,
mirrors the IQR with the best score for $u_{SIP}$ and $u_{Cox}$
and the worst for $u_{NIP}$.

For the large sample ($N=40$), The trends noted above are still perceptible,
but at a lesser level. The three location estimators give more consistent
results. $u_{SIP}$ stay significantly better than the other estimators
for the $P_{20}$ statistic, followed by $u_{MIP}$.

\subsection{Comparison of uncertainty estimators for non-normal distributions
of errors}

\begin{figure}[t]
\noindent \begin{centering}
\includegraphics[height=7cm]{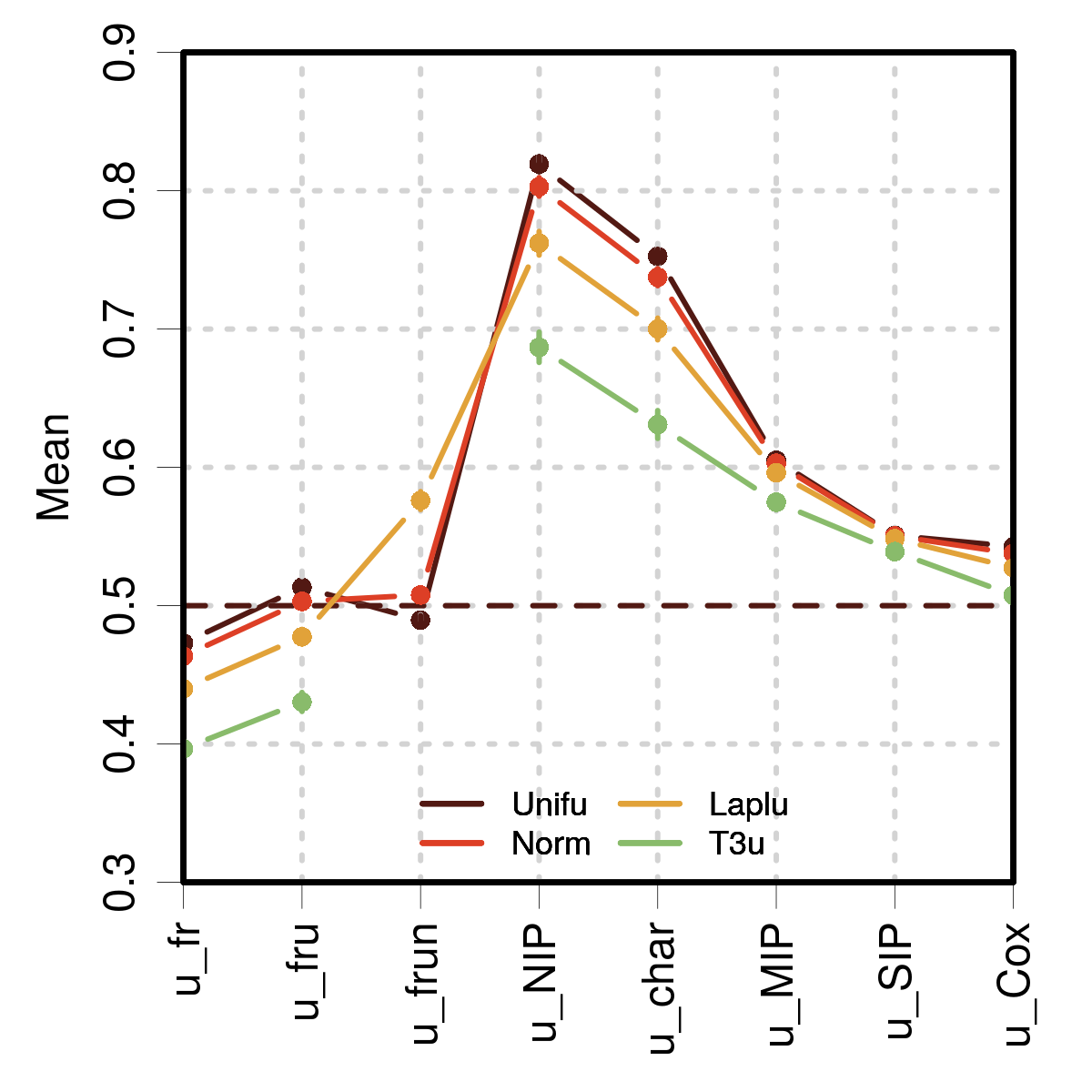}\includegraphics[height=7cm]{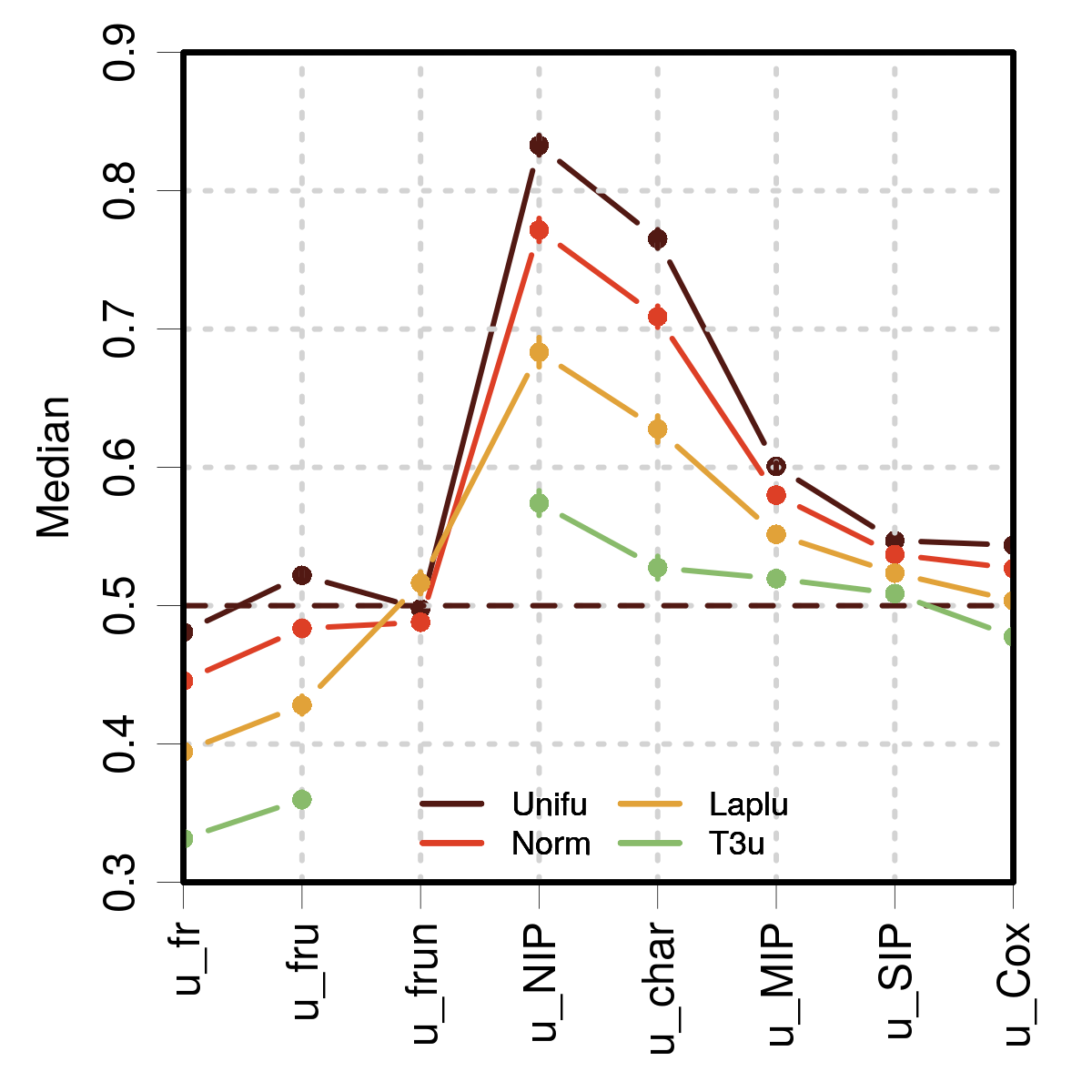}
\par\end{centering}
\noindent \begin{centering}
\includegraphics[height=7cm]{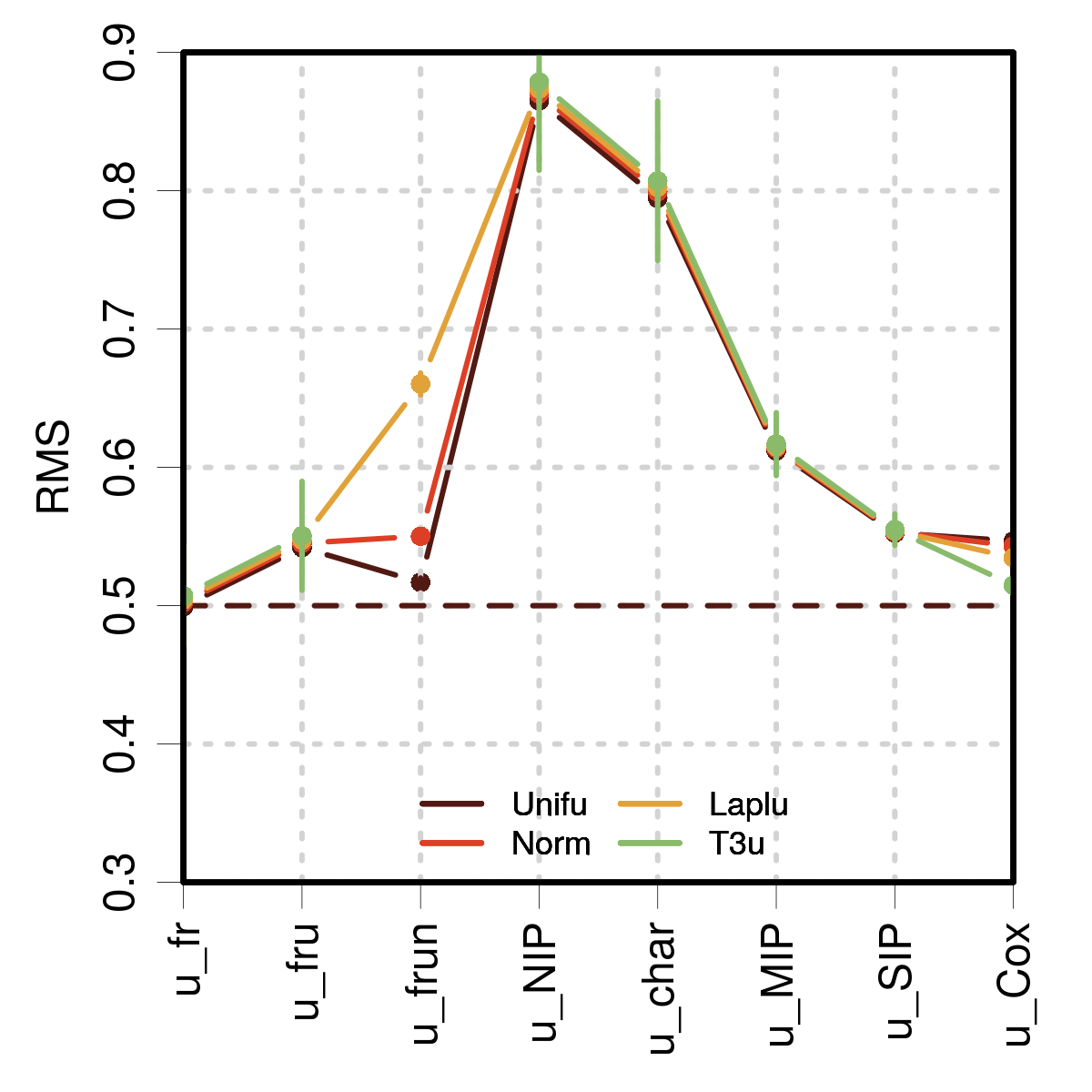}\includegraphics[height=7cm]{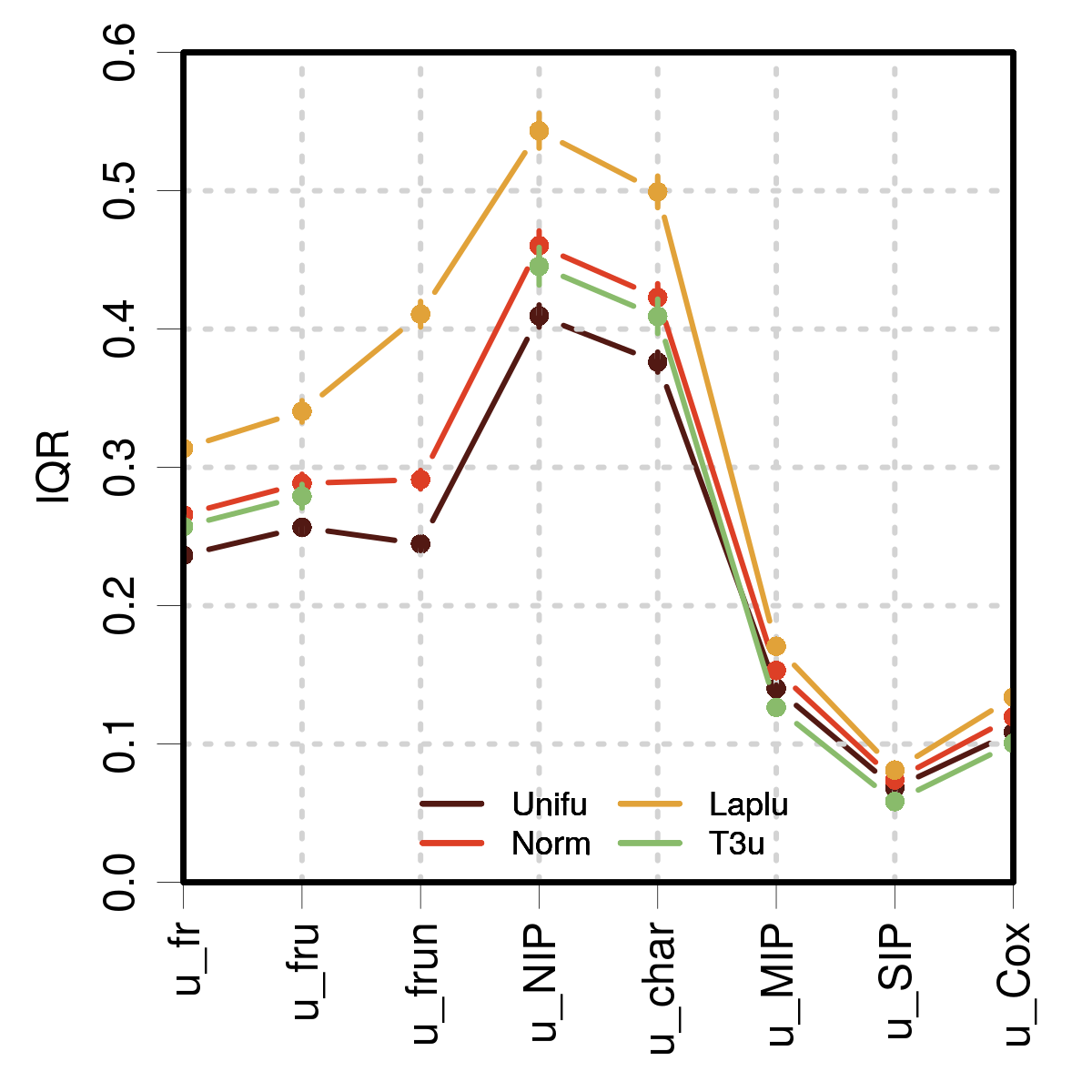}
\par\end{centering}
\noindent \begin{centering}
\includegraphics[height=7cm]{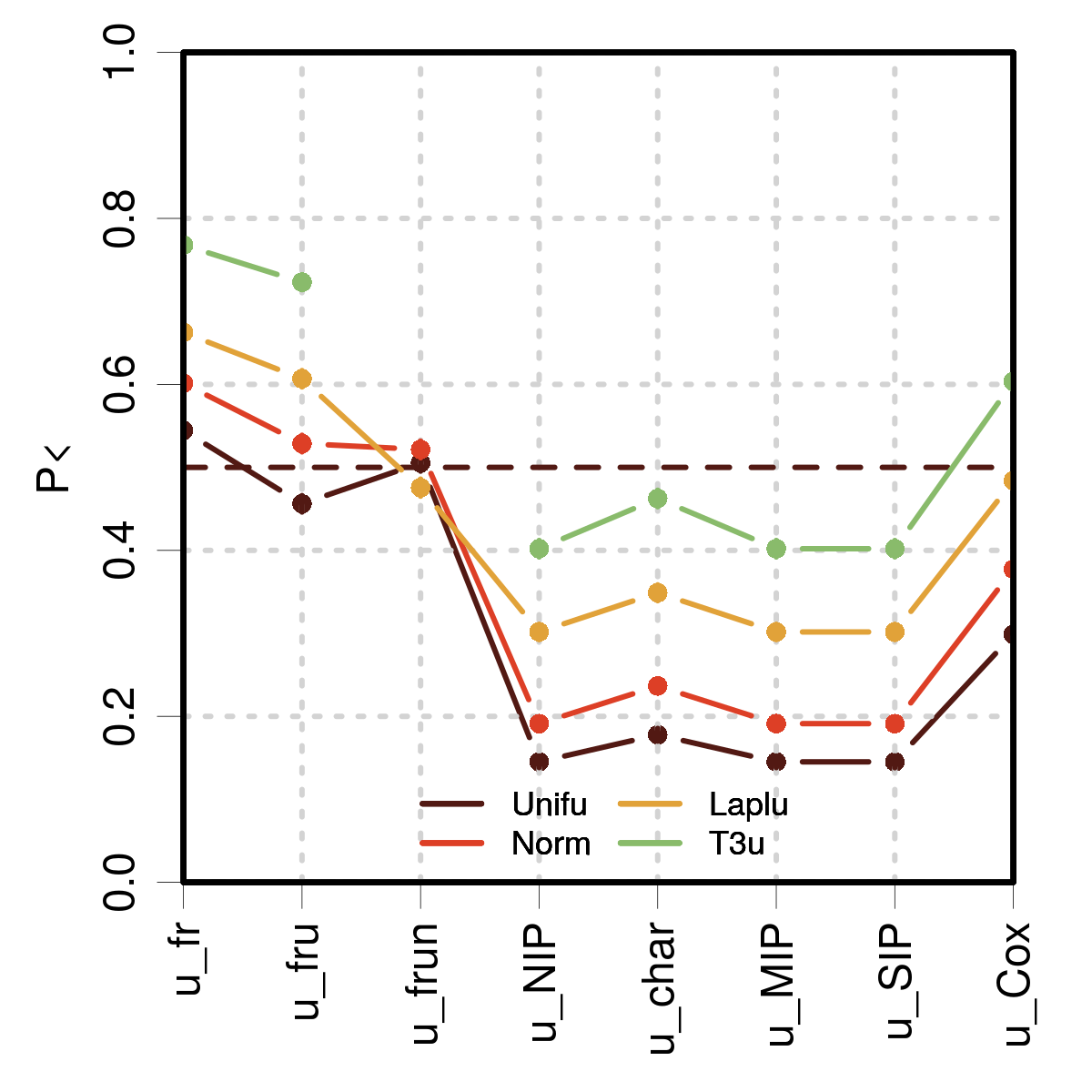}\includegraphics[height=7cm]{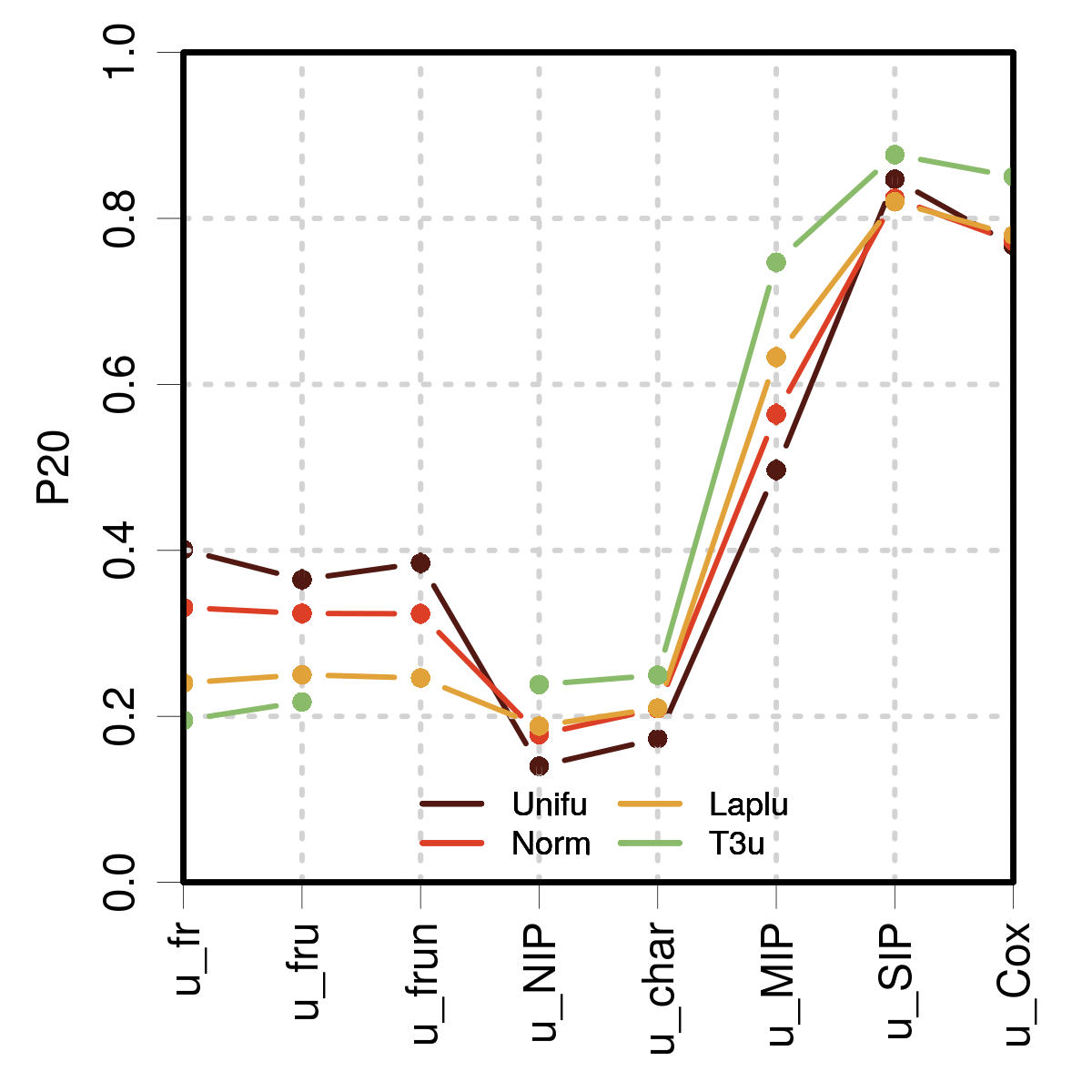}
\par\end{centering}
\caption{\label{fig:Statistics4}Statistics for $N=4$.}
\end{figure}

\begin{figure}[t]
\noindent \begin{centering}
\includegraphics[height=7cm]{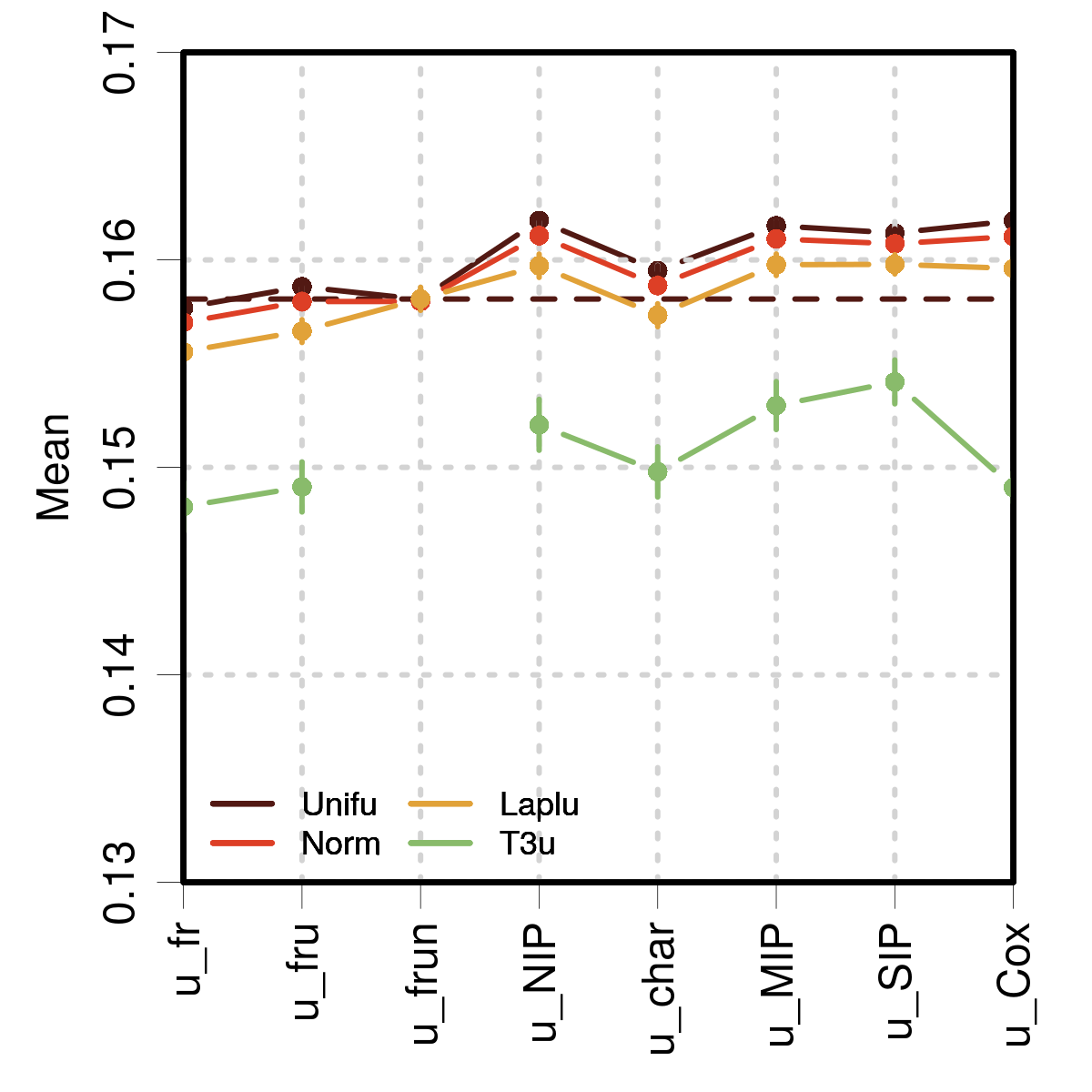}\includegraphics[height=7cm]{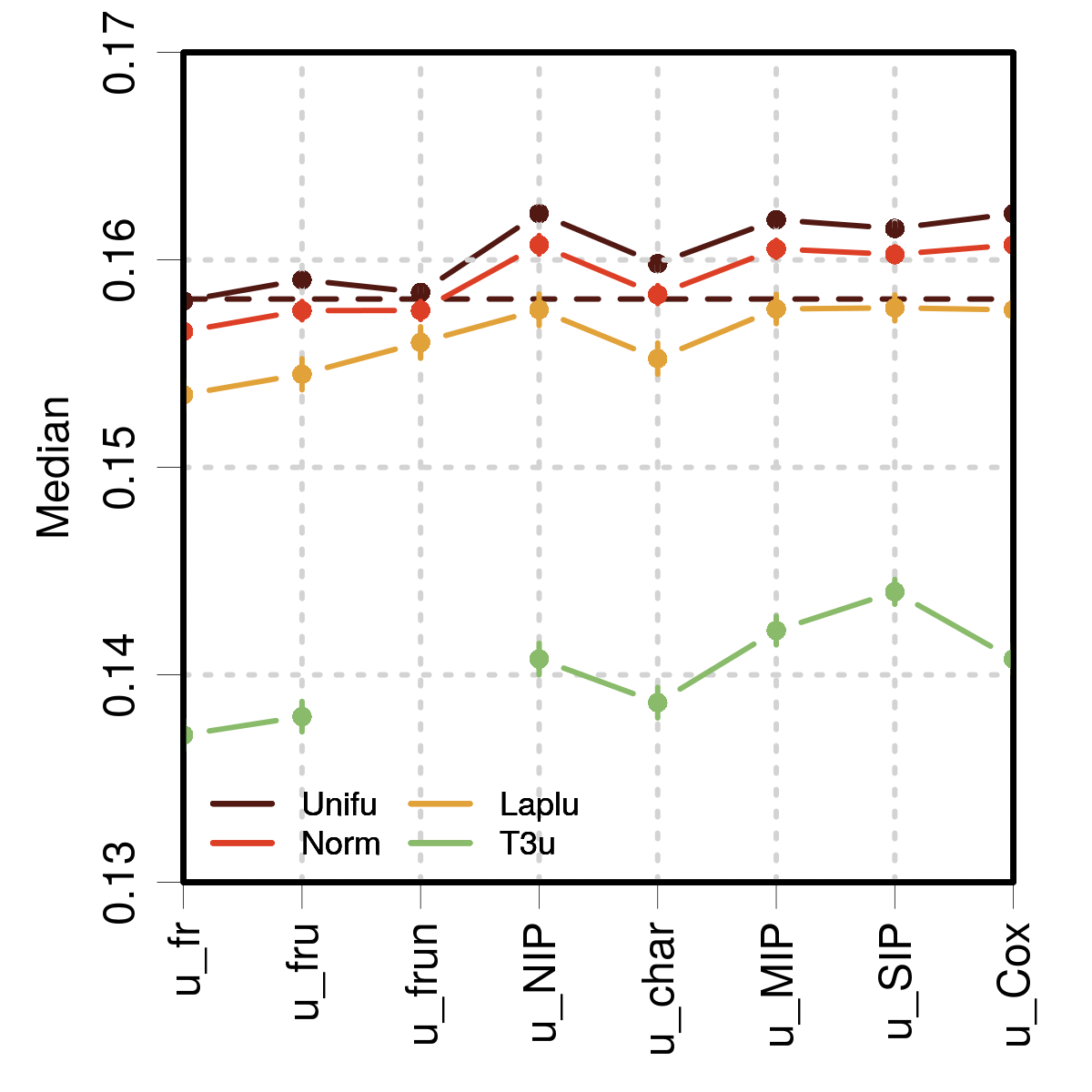}
\par\end{centering}
\noindent \begin{centering}
\includegraphics[height=7cm]{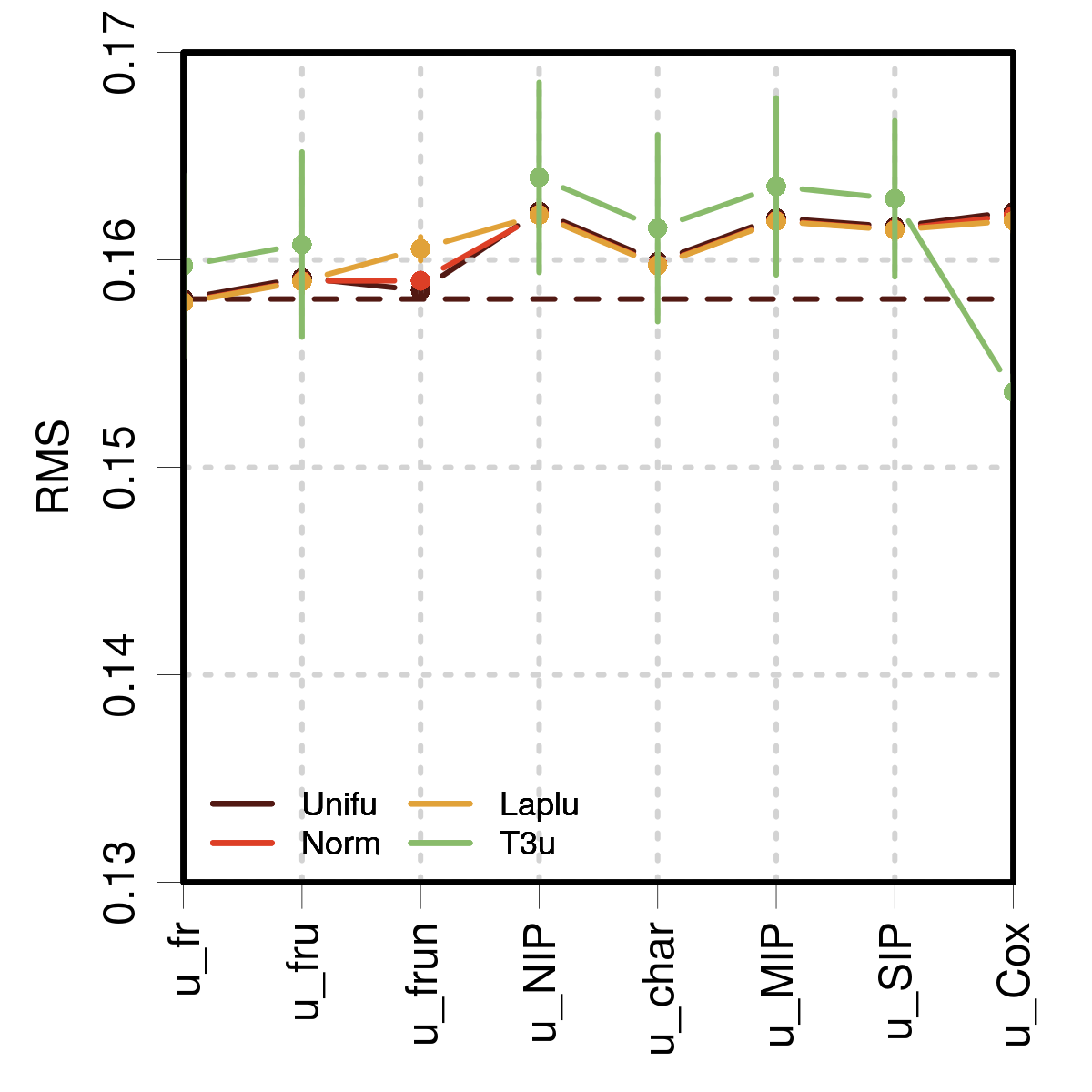}\includegraphics[height=7cm]{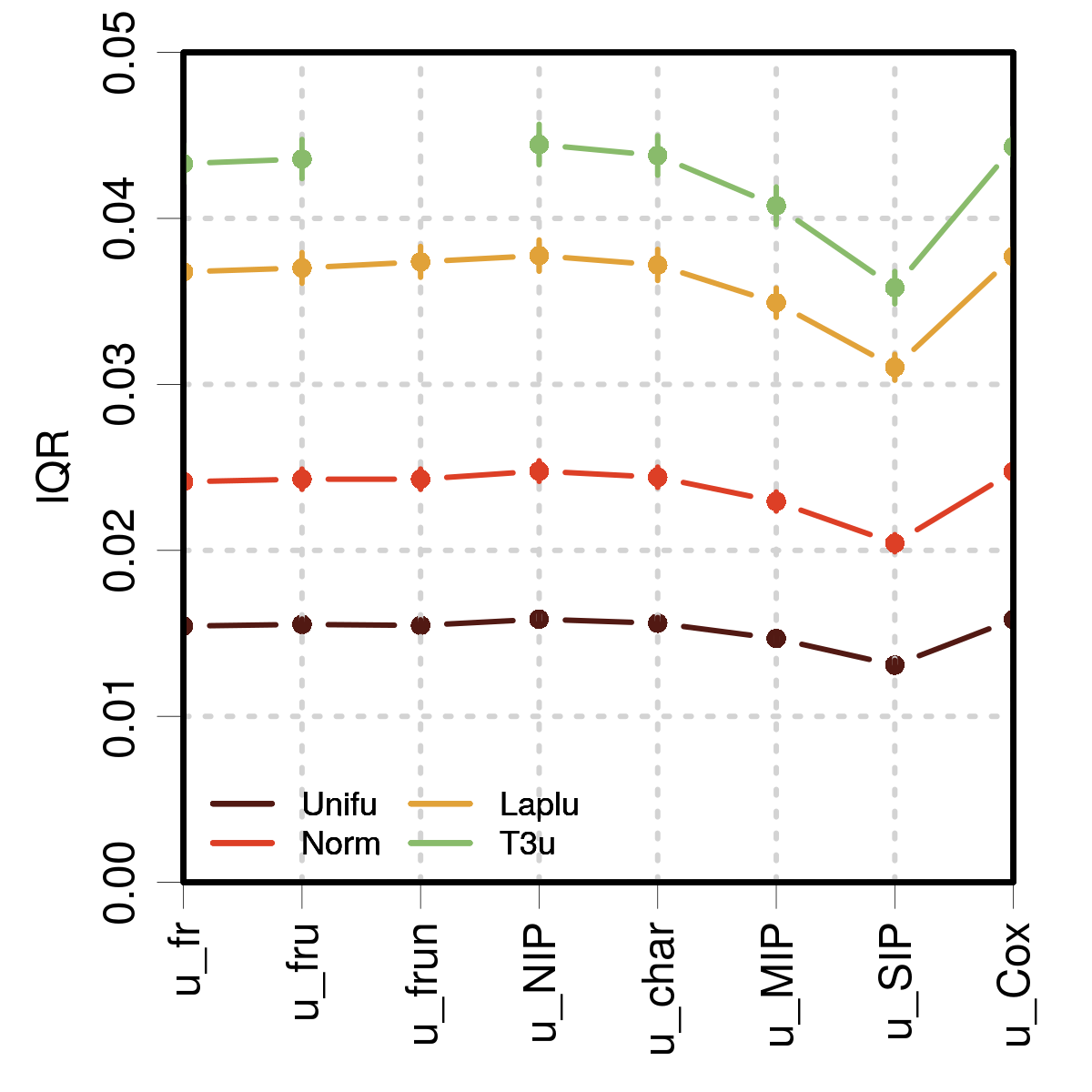}
\par\end{centering}
\noindent \begin{centering}
\includegraphics[height=7cm]{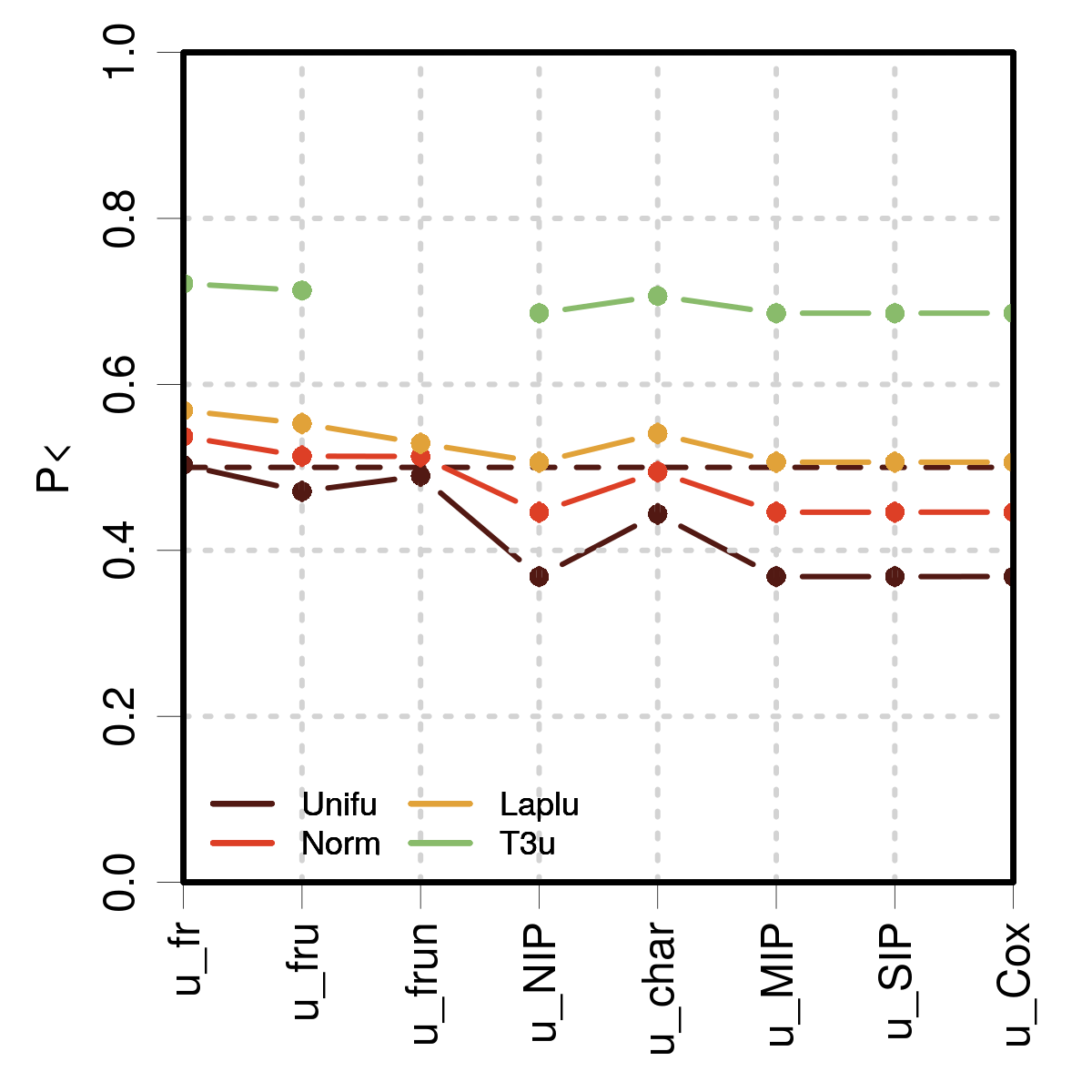}\includegraphics[height=7cm]{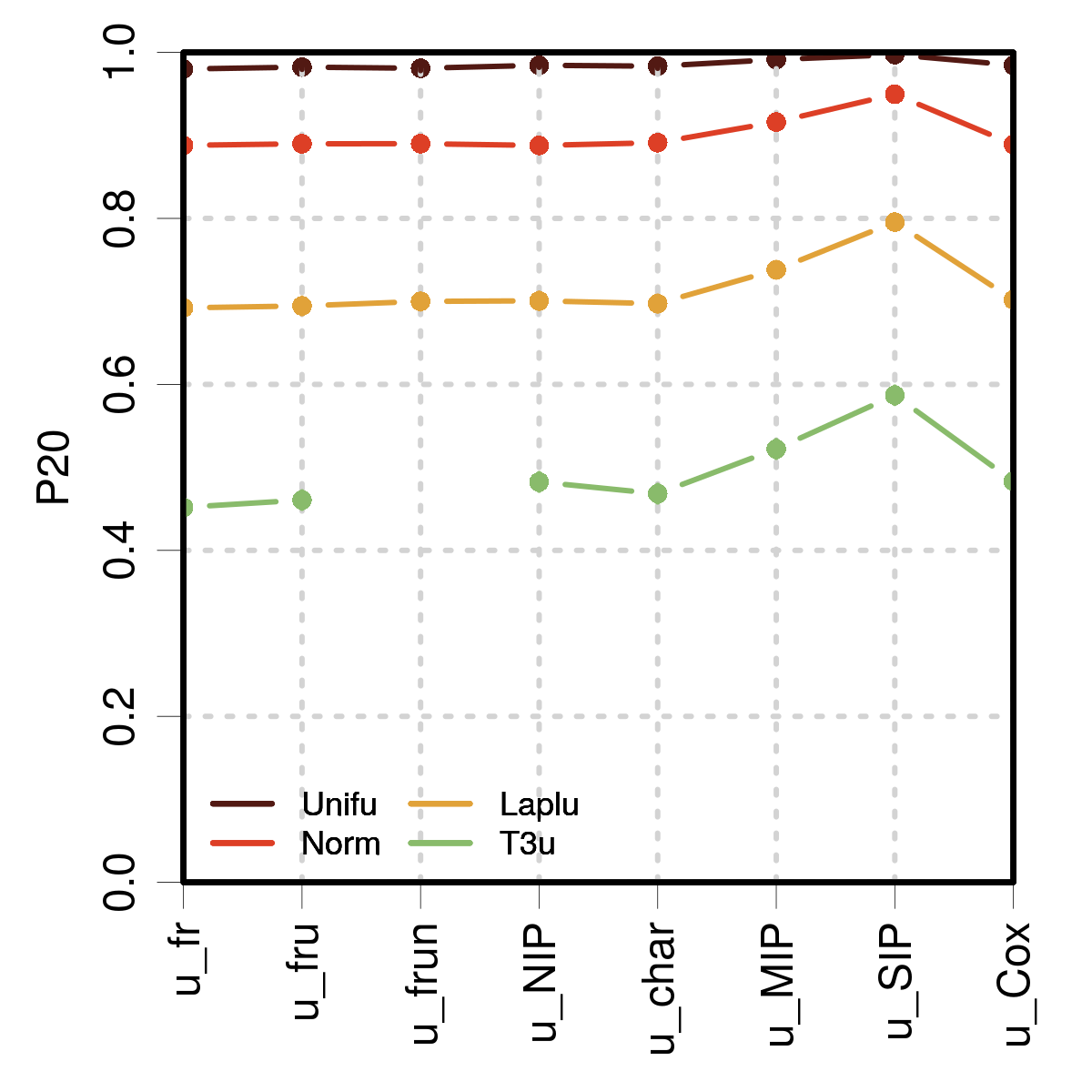}
\par\end{centering}
\caption{\label{fig:Statistics40}Statistics for $N=40$}
\end{figure}

The same statistics as above have been calculated for non-normal distributions.
The results are reported in Figs.\,\ref{fig:Statistics4} for $N=4$
and Fig.\,\ref{fig:Statistics40} for $N=40$. 

Several trends are clearly visible in the plots summarizing the statistics
for $N=4$. For instance, the base frequentist formula $u_{fr}$ consistently
underestimates the standard uncertainty, except when using the RMS
estimation. It should be noted also that the RMS estimation of the
mean uncertainty is the least sensitive to the non-normality of the
distribution (except for $u_{frun)}$), the most sensitive one being
the median. The $u_{frun}$ estimator, accounting for the excess kurtosis
of the distributions seems to perform better for the median than for
the mean or RMS. As a consequence, it is also the less sensitive to
kurtosis when estimating $P_{<}$. However, it is not able to deal
with T3u, which has an infinite kurtosis. The unbiased estimator $u_{fru}$
performs better (by design) for the normal distribution (Norm), but
also for the uniform distributions (Unifu). However, it fails for
the Student's distribution (T3u), and to a lesser extent for the Laplace
distribution (Laplu). 

If one ignores the $u_{frun}$ estimator, one can note a systematic
effect of kurtosis (all curves are parallel), and a global difficulty
to deal with the T3u distribution, except for the RMS metric which
presents a small offset but larger statistical uncertainties.

The performance of the Bayesian estimators depends notably on the
information level introduced by the prior:
\begin{itemize}
\item the non-informative $u_{NIP}$, mildly informative $u_{MIP}$ and
strongly informative $u_{SIP}$ priors result in consistent over-estimation
of the standard uncertainty, for either Mean, Median or RMS;
\item for all statistics, the $u_{char}$ estimator is closer to the target
than $u_{NIP}$, although not better than the more informed Bayesian
estimators;
\item by comparison, the Cox prior $u_{Cox}$ provides about the same results
than $u_{SIP}$, but is more sensitive to the kurtosis of the distribution
for the three location statistics.
\end{itemize}
The IQR is also sensitive to kurtosis, but notably less for the Bayesian
estimators based on informative priors, which provide more concentrated
distributions, as noted above for the normal case. This is reflected
in $P_{20}$ for which $u_{SIP}$ and $u_{Cox}$ reach levels near
0.8.

For the larger sampling size $N=40$, the discrepancies between the
estimators are strongly reduced. Concerning the location metrics,
the mean is still sensitive to kurtosis, except for $u_{frun}$, the
median is now the most sensitive to kurtosis, even for $u_{frun}$,
and except for the T3U distribution and $u_{frun}$ estimator, the
RMS metric is still the less sensitive to kurtosis.

All the Bayesian estimators are still on the side of overestimation
and the impact of the prior is less sensible than for $N=4$. Surprisingly,
$u_{char}$ seems to perform better than all Bayesian estimators,
at least for the Unifu and Norm distributions.

IQR and $P_{20}$ are still notably sensitive to kurtosis. The best
performers for all distributions are now $u_{MIP}$ and $u_{SIP}$.
$u_{Cox}$ has lost the advantage it had for the small sample. Even
for sets of 40 measurements, the impact of the additional information
introduced by the MIP and SIP priors is non-negligible. The downside
is that a wrongly designed prior might need a lot of data to be countered.
\begin{verbatim}

\end{verbatim}

\section{Conclusion\label{sec:Conclusion}}

Our Monte Carlo simulations enable to dress a contrasted portrait
on the mean uncertainty estimation landscape. To evaluate the performances
of a comprehensive set of uncertainty estimators, several statistics
were used to summarize the properties of the uncertainty distribution
generated by Monte Carlo sampling. Three location statistics (mean,
median, RMS), a spread statistic (IQR) and two probabilities, $P_{<}$
the probability of underestimation and $P_{20}$ the probability to
lie within 20\% of the true value. Several location statistics were
used to account for the skewness of the uncertainty distributions,
notably for small measurement samples.

For small samples, $u_{fr}$ provides closer estimates when combined
with RMS, while $u_{fru}$ works better with the mean. As evoked earlier,
this is consistent with the fact that $u_{fr}$ derives from an unbiased
estimator of variance, while $u_{fru}$ is an unbiased estimator of
standard deviation. As the GUM is by essence built on the rule of
combination of variances, one might question if an unbiased estimator
of standard deviation is pertinent for metrology. Additionally, we
have seen that the RMS statistic is much less dependent on the kurtosis
of the error distribution than the other location statistics. One
might thus base the discussion on the RMS results. 

In this case, $u_{fr}$ is the uncertainty estimator which gives the
best estimate in the long run. By contrast, all other estimators tend
to overestimate uncertainty. For instance, all tested Bayesian estimators
give a notable overestimation of the uncertainty on the mean, which
is better corrected by Cox's informative prior and the strongly informative
prior of O'Hagan and Cox. 

A remarkable point is that the MIP, SIP and Cox estimators give an
uncertainty distribution that is much more concentrated than the other
ones, therefore a lesser risk of strong under- or over- estimation.
This comes however with a caveat about the necessity of a good prior
parameterization.

Globally, there is a significant gain in using the Bayesian estimators
with informative priors. For small measurement sets, they provide
more conservative uncertainty values, but also a very good probability
to lie within some tight interval around the true value. On all aspects,
the estimator based on a non-informative prior $u_{NIP}$ does not
seem to be a good choice, even worse than the frequentist formulas.

Finally, one should be aware that all estimators, even the unbiased
ones, are at some level sensitive to the shape of the error distribution,
and seem to be at pain for distributions with strong positive excess
kurtosis {[}Laplace or Student's-\emph{t}($\nu=3$){]}. 

\section*{Data availability statement }

The \texttt{R} code that enables to reproduce the figures and tables
of this study is openly available at the following URL: \url{https://github.com/ppernot/2022_SampleMean},
or in Zenodo at \url{https://doi.org/10.5281/zenodo.7063942}.

\bibliographystyle{unsrturlPP}
\bibliography{NN}

\end{document}